\documentclass[a4paper,twoside]{article}
\newcommand{\affil}[1]{$^{\rm #1}$}

\usepackage[authoryear]{natbib}
\bibpunct{(}{)}{;}{a}{}{,}
\usepackage{graphicx}

\date{}

\begin{document}

\title{PHOTZIP: A Lossy FITS Image Compression Algorithm that Protects User-Defined Levels of Photometric Integrity}

\author{\parbox{\textwidth}{\flushleft
\vspace{-0.5cm}
{\it Lior Shamir\affil{A}, Robert J. Nemiroff\affil{A}}\\
\vspace{0.4cm}
{\small \affil{A}\,Michigan Technological University, Department of Physics \\ 1400 Townsend Drive, Houghton, MI  49931}\\
}}

\vspace{0.4cm}

\twocolumn[
\begin{changemargin}{.8cm}{.5cm}
\begin{minipage}{.9\textwidth}
\vspace{-1cm}
\maketitle

\small{\bf Abstract:}
A lossy compression algorithm is presented for astronomical images that protects photometric integrity for detected point sources at a user-defined level of statistical tolerance.  PHOTZIP works by modeling, smoothing, and then compressing the astronomical background behind self-detected point sources, while completely preserving values in and around those sources.  The algorithm also guaranties a maximum absolute difference (in terms of $\sigma$) between each compressed and original background pixel, allowing users to control quality and lossiness.  For present purposes, PHOTZIP has been tailored to FITS format and is freely available over the web. PHOTOZIP has been tested over a broad range of astronomical imagery and is in routine use by the Night Sky Live (NSL) project for compression of all-sky FITS images.  Compression factors depend on source densities, but for the canonical NSL implementation, a PHOTZIP (and subsequently GZIP or BZIP2) compressed file is typically 20\% of its uncompressed size.

\medskip{\bf Keywords:} methods: data analysis -- techniques: image processing, photometric -- astronomical data bases: miscellaneous

\medskip
\medskip
\end{minipage}
\end{changemargin}
]
\small

\section{Introduction}

Astronomy research images are now almost exclusively digital.  There is an increasing need to store these images and to send them over the Internet.  Average bandwidth and storage capacity, although increasing, continue to be a bottleneck that frequently limits scientific exploitation of these images.  Lossless compression of astronomical image files therefore creates a clear advantage over non-compression since it increases effective storage and bandwidth and so bolsters scientific utility.  Lossy compression is more controversial, however, as the scientific value of the data lost in the compressed images must be weighed against the scientific value of the data gained by the extra bandwidth and storage space. 

One of the premier scientific uses for astronomical images is photometry.  Whether detecting the presence of, for example, distant supernovae, Local Group microlensing, binary star variability, or planetary transits, the need for photometric accuracy in astronomy images remains a primary objective for many astronomical research projects.

Frequently, the photometric value of an astronomical image is concentrated in the point sources in the image.  Conversely, the bulk of the image size is concentrated in the background behind these sources.  When the number of pixels taken up by sources is small compared to the number of pixels that compose the background, it becomes possible to significantly compress the image size while preserving a certain level of photometric integrity.

While lossless compression algorithms preserve 100\% of the signal, lossy algorithms can provide a better compression factor while losing some of the signal \citep{14,6,7}. However, lossy compression algorithms tend to convolve science with art.  Astronomy-specific lossy data compression algorithms are not new, and many have been proposed and widely used \citep{3,4,5,13,14}. HCOMPRESS \citep{13} is a commonly used FITS compression program based on a two-dimentional Haar wavelet transform. This compression program is fast and provides a relatively high compression factor. However, while providing the user control of the lossiness$\backslash$compression tradeoff, HCOMPRESS provides a limited control over the type of the signal that is lost in the compression $\backslash$ decompression process.

The most similar compression approach to that discussed here is the insightful FITS compression program FITSPRESS \citep{14}. FITSPRESS is a wavelet based compression algorithm that has, among other things, sensitivity to preserving the brightest image pixels.

Lossy compression (e.g. JPG, HCOMPRESS) tends to be controlled by parameters that have no scientific or statistical meaning.  In contrast, the PHOTZIP FITS compression algorithm provides the user control of the preserved$\backslash$lost signal in scientific terms. Like \citep{18}, PHOTZIP should be used as a preprocessor to multi-purpose lossless compression algorithms (e.g. LZW) and allows those algorithms a higher compression factor by losing some of the signal, but ensures that only {\it background} information of the image will lose some of its signal. The algorithm provides an interface which can be used in order to define, in terms of $\sigma$, the criteria for a pixel to be considered as background, and guarantees a user-defined maximum absolute difference (also in terms of $\sigma$) for any background pixel that loses signal.  In Section 2 we describe the algorithm, in Section 3 we present ways to improve the compression factor, in Section 4 we discuss photometric integrity and in Section 5 we discuss the performance of the algorithm.

\section{Lossy Compression That Preserves Bright Signals}

Since a main purpose of our algorithm is to allow lossiness only for background pixels, the first stage of the algorithm is to determine for every pixel in the frame whether or not it is a background pixel.  For PHOTZIP, we achieve this by using square window median filtering.  The background value of each pixel is determined to be the median of the values of all pixels within its window. Assuming that the bias is zero, the gain is 1 electron and the read noise is negligible, we can compute $\sigma$ for each pixel by $\sigma = \sqrt{B_{x,y}}$, where $B_{x,y}$ is the estimated background of the pixel at coordinates $(x,y)$. \newline
The background computation stage can be summarized by the following algorithm: \newline \newline
1. for y $\leftarrow$ 1 to height do \newline 
2. $\indent$ for x $\leftarrow$ 1 to width do \newline
3. $\indent \indent B_{x,y} \leftarrow$ median of $C_{x-s,y-s}, C_{x-s+1, y-s},$ \newline $ \ldots , C_{x-s,y-s+1}, \ldots ,C_{x+s,y+s}$ \newline
4. $\indent \indent \sigma_{x,y} = \sqrt{B_{x,y}}$ \newline
5. $\indent$ end for \newline
6. end for \newline \newline 
$C_{x,y}$ is the value of the pixel at the coordinates $(x,y)$, and $B_{x,y}$ is the estimated background value of the pixel at the coordinates $(x,y)$. $s$ is the half width size of the window. The nested loops in lines 1 and 2 make sure that the background computation is done for every pixel.  Therefore, every pixel in the frame is attached to the background value that is the median of the pixels in the $(2s + 1) \times ( 2s + 1)$ window centered on $(x,y)$. After the estimated background is computed for a pixel, the $\sigma$ of that pixel is also computed by $\sigma = \sqrt{B_{x,y}}$.  In our current implementation, PHOTZIP assumes a default half-width size of 10, although the user can specify any half-width size.

When an image has a great many pixels, a pixel-by-pixel median computation might turn into a computationally expensive task. For an $N \times N$ frame we will need to compute the median value of $N^2$ square windows, when each window has $(2s+1) \times (2s+1)$ pixels. For instance, for a 1024 $\times$ 1024 frame, when computing the background values of the pixels using a windows with half width size of 10, The algorithm will need to compute 1,048,576 (1024 $\times$ 1024) times a median of 441 (21 $\times$ 21) numbers. In order to reduce needed processing power, we chose to compute the background value not for {\it every} pixel, but for windows of 5 $\times$ 5 in which the median value is computed only for the ``leader" pixel, which is the pixel at the center of the window. All other pixels in that 5 $\times$ 5 window are associated with the same background value as their ``leader". Since backgrounds do not tend to drastically change over small windows, we consider this technique as an acceptable approximation. In order to compute the median efficiently, we use the common algorithm for finding a median in linear time described by \citep{10}.

Once the background value (and hence $\sigma$) is determined for every pixel in the frame, the pixel values are quantized. The quantization stage is affected by two parameters.  The first, $d$, is the minimum brightness, in terms of $\sigma$, of a pixel such that every pixel that is less bright is classified as background.  The second parameter, $b$, is the maximum absolute difference, also in terms of $\sigma$, that is allowed between a background pixel in the original image and the same pixel in the compressed$\backslash$decompressed image. The basic idea of the quantization is that a value of a certain pixel $(x,y)$ is quantized only if it is lower than $B_{x,y}+ d\cdot\sigma_{x,y}$, so the lossiness of the algorithm does not affect sufficiently bright pixels. For every pixel $(x,y)$ in the frame, we first check if it is brighter than $B_{x,y} + d\cdot\sigma_{x,y}$. If the pixel meets this criterion then it is not quantized and does not loose any of its signal.

The criteria for a pixel to be quantized is 
\begin{equation}
 C_{x,y} < B_{x,y} + d \cdot \sigma_{x,y} .
\end{equation} 
The top-level algorithm for selecting the pixels that should be quantized is:\newline \newline 
1. for y $\leftarrow$ 1 to height do \newline 
2. $\indent$ for x $\leftarrow$ 1 to width do \newline
3. \indent \indent if $C_{x,y}-B_{x,y} < d \cdot \sigma_{x,y}$ then \newline
4. \indent \indent \indent $C_{x,y} \leftarrow quantize(C_{x,y},\sigma_{x,y},b)$ \newline
5. $\indent$ end for \newline
6. end for \newline \newline
In line 4, the subroutine {\it quantize} is called in order to perform the quantization of any pixel that does not meet the criteria of line 3. The size of the quanta used is $2 \cdot {2^{ \lfloor \log _{2} b \cdot \sigma \rfloor}}$, where $b$ is a user-defined positive value ($b>0$) such that the maximum absolute difference between the value of the pixel in the original frame, and the value of the same pixel in the compressed$\backslash$decompressed frame cannot be greater than $2^{ \lfloor \log _{2} b \cdot \sigma \rfloor}$.
\newline
The quantization algorithm is simply: \newline \newline 
quantize ($c$,$\sigma$,$b$) \newline
1. $quantum\_size \leftarrow 2 \cdot {2^{ \lfloor \log _{2} b \cdot \sigma \rfloor }}$ \newline
2. $quantized\_value \leftarrow$ $quantum\_size \cdot Round({\frac{c}{ quantum\_size }})$ \newline
3. return($quantized\_value$) \newline
\newline
$c$ is the pixel's value and $b$ is the maximum absolute difference (in terms of $\sigma$) that is allowed between a pixel in the original frame and the same pixel in the compressed$\backslash$decompressed frame. {\it Round} is a function that rounds its argument to the nearest integer. This quantization symmetrically increases or decreases pixel values. The interval of each quantum is $[quantized\_value-{2^{ \lfloor \log _{2} b \cdot \sigma \rfloor }}  ,quantized\_value+{2^{ \lfloor \log _{2} b \cdot \sigma \rfloor }}]$. Since $c$ is a value within this interval, the absolute difference $|c-quantized\_value|$ can never be greater than ${2^{ \lfloor \log _{2} b \cdot \sigma \rfloor }}$. Since ${2^{ \lfloor \log _{2} b \cdot \sigma \rfloor }} \leq b\sigma$, the absolute difference between a pixel value in the compressed$\backslash$decompressed frame and the same pixel in the original frame can not be greater than $b\sigma$. It might seem that simple quanta at the size of $2 b \sigma$ can improve the compression factor by providing larger quantum sizes, yet still comply with the maximum absolute difference criteria. However, since $\sigma$ is different for every pixel, this might lead to a large variance in the quantum sizes and therefore severely reduce the compression factor. For instance, suppose that we have two pixels with values of 97 and 99, and $\sigma$ of 9 and 10 respectively. Assuming $b=1$, with quantum size of $2b\sigma$ the first value, 97, will be quantized using quantum sizes of 18 and the second value, 99, will be quantized using quantum sizes of 20. After the quantization process, the values will be, therefore, 90 and 100 respectively. However, if using quantum size of ${2^{ \lfloor \log _{2} b \cdot \sigma \rfloor + 1}}$, the quantization of both values is done using the same quantum size (16 in this case). After the quantization process, the value of both pixels will be 96, which increases the compression potential of pattern matching based compression algorithms. The low variance of quantum sizes leads to smaller variance of quantized values, which is an important factor in the performance of many multi-purpose compression algorithms. Examples of the differences in the compression factor (using {\it BZIP2} with PHOTZIP) when using quantum size of $2b\sigma$ and quantum size of $2 \cdot {2^{ \lfloor \log _{2} b \cdot \sigma \rfloor }}$ are listed in the following table:

\def\0{\phantom{0}}
\halign{#\hfil& \quad#\hfil& \quad#\hfil& \quad#\hfil& \quad#\hfil& \quad#\hfil& \quad#\hfil\cr
$d$& $b$& $s$& \bf quantum size & quantum size \hfill\cr
& & & \bf $2 \cdot {2^{ \lfloor \log _{2} b \cdot \sigma \rfloor }}$& $2b\sigma$ \hfill\cr

1& 1& 8& 73.6\%& 64.2\%\cr
2& 1& 8& 76.5\%& 66.1\%\cr
1& 2& 10& 77.3\%& 67.0\%\cr
}

The file that was used for the samples is ``ci040325ut 005115p.fits" which is an unsigned integer FITS image of size of 2102 KB. This file is discussed more thoroughly in Section 5.

The function {\it quantize} might fail when $\sigma$ is equal to zero. However, when $\sigma=0$, the condition stated in line 3 of the top level algorithm cannot be satisfied and the function {\it quantize} is not invoked.

Since the quantization is symmetric, the mean of the pixel values should be preserved in the compressed $\backslash$ decompressed frame. Assuming a normal distribution for the pixel values in the original frame, the median of the original frame should be equal to the mean. As the mean is preserved, the median of the original frame can be taken from the mean of the compressed $\backslash$ decompressed frame.

Since the number of integers within each \newline $[quantized \_value - {2^{ \lfloor \log _{2} b \cdot \sigma \rfloor }},quantized\_value+{2^{ \lfloor \log _{ 2} b \cdot \sigma \rfloor }}]$ quantum is odd, there is always one integer value that can be quantized either up or down. A systematic policy that always quantizes these values in the same fashion (either up or down) will cause a systematic bias to the mean. Thus, in order to avoid statistical biasing, the standard {\it Round} function always rounds half integers to the nearest even integer \citep{19}. Due to that behavior of {\it Round}, if the quantized value is exactly in the center of the interval, it can either gain or lose value so that systematic bias is avoided.

\section{Compressing Non-Astronomical Edges and Artifacts}

In practice, image pixels might exist that are clearly non-astronomical and do not need to be preserved.  One example is the edge of a frame that is not exposed to the sky but dominated by other sources of noise.  In some cases, pixels such as these would not have a high pixel value, but since they are in a relatively dim area of the frame, their value is high compared to their background. Pixel values of non-astronomical edges can sometimes have a high variance, so some of the pixels can be significantly brighter than other pixels around them. Since $\sigma$ is determined by the local background of each pixel, a low background value of a pixel leads to a low $\sigma$, and a low $\sigma$ leads to a low $d\sigma$. Given their low $d \sigma$, the algorithm would normally preserve the signal of pixels which do not have a high value by meeting the $C_{x,y} > B_{x,y} + \sigma_{x,y} d$ criteria due to the extremely low value of their background. This might result in an unnecessarily low compression factor. In order to allow a user to avoid this trap, we set another optional parameter $t$ that sets a threshold value for pixels that are allowed to lose signal. The threshold value is set by $t$ in terms of the median value of the frame. For instance, if $t=1.5$ then any pixel with value less than $1{1 \over 2}$ times the value of the median pixel will not be required to preserve its value even if it meets the $C_{x,y} > B_{x,y} + \sigma_{x,y} d$ criteria. Therefore, to include this, line 3 of the top level algorithm presented in section 2 should be changed to: \newline
3. \indent if $C_{x,y}-B_{x,y} < d \cdot \sigma_{x,y}$ or $C_{x,y} < t \cdot $ 
{\rm (median of} ${C_{1,1}, C_{2,1}, \ldots , C_{1,2}, \ldots ,C_{width,height}})$ then \newline
\newline
We found this parameter is very effective in compressing Night Sky Live \citep{1} FITS frames, in which a significant portion of the frame is not directly exposed to the sky, and therefore consists of scattered low values that have no scientific utility.  If those areas were uniform, then no special action would need to be taken and the pixel values would be naturally quantized.

\section{Photometric Integrity}

Many astronomical images are sparsely populated.  When the number of pixels taken up by point sources is small compared to the number of pixels that compose the background, it becomes possible to significantly compress the image while preserving a useful level of photometric integrity.  In this Section we estimate this level of photometric integrity.

When symmetrically quantizing the background pixels, the mean of the pixel values is preserved in a level that depends on the number of background pixels averaged.  A high number of averaged pixels will lead to a low standard error. 

When a pixel value is quantized, the absolute difference between the pixel value in the original frame and the pixel value in the compressed$\backslash$decompressed frame is bound by ${2^{ \lfloor \log _{2} b \cdot \sigma \rfloor }}$. Let $\Delta$ be the difference between the value of a certain pixel $(x,y)$ in the original frame and the value of the same pixel in the compressed$\backslash$decompressed frame such that: \newline
$\Delta = C_{x,y} - $quantize$(C_{x,y},\sigma,b)$ \newline
Let $\bar C$ be the mean of $n$ pixel values in the compressed $\backslash$ decompressed frame such that: \newline
$\bar C = {\frac {\sum_{i=1}^{n}{(C_{i}-\Delta_{i})}}{n}} = {\frac {\sum_{i=1}^{n}{C_{i}}}{n}} - {\frac {\sum_{i=1}^{n}{\Delta_{i}}}{n}}$ \newline
So $|{\frac {\sum_{i=1}^{n}{\Delta_{i}}}{n}}|$ is the absolute difference between the mean of the original pixel values and the mean of the quantized pixel values. \newline
Let $q={2^{ \lfloor \log _{2} b \cdot \sigma \rfloor }}$.
Since the function {\it quantize} described in section 2 guaranties that the absolute difference $|\Delta|$ is bound by $q$, $\Delta$ can be any value within the interval $[- q, q]$.  Since $\Delta$ is uniformly distributed in the interval $[- q,q]$, the expected value of $\Delta$ is zero, and the standard deviation of $\Delta$ is ${q \over {\sqrt{3}}}$. \newline
$stddev(\Delta)= {q \over {\sqrt 3}}$ \indent ({\it stddev(X)} is defined as the standard deviation of a random variable X). \newline
The variance of $\Delta$ is:\newline
$var(\Delta) = stddev^{2}(\Delta) = {\frac {q^{2}}{3}}$ \newline
Let $\bar \Delta$ be the absolute differences between the mean of the original values and mean of the quantized values of some $n$ pixels.\newline
$\bar \Delta = {\frac {\sum_{i=1}^{n} \Delta_{i}}{n}} = {\sum_{i=1}^{n}{\frac {\Delta_{i}}{n}}}$ \newline
$var(\bar \Delta) = var({\sum_{i=1}^{n}{\frac {\Delta_{i}}{n}}}) = {\sum_{i=1}^{n}{var({\Delta_{i} \over n})}} = {\sum_{i=1}^{n}{\frac {var(\Delta_{i})}{n^{2}}}} = {\frac {n \cdot var(\Delta)}{n^{2}}}={\frac {var(\Delta)}{n}}$ \newline
$stddev(\bar \Delta) = {\sqrt {var(\bar \Delta)}} = {\sqrt {\frac {var(\Delta)}{n}}} = {\sqrt {\frac {q^{2}}{3n}}} = {{q} \over {\sqrt {3n}}} = {{2^{ \lfloor \log _{2} b \cdot \sigma \rfloor }} \over {\sqrt {3n}}}$
\newline
\newline
We can see that the most likely value of the difference between the mean of the original values and the mean of the quantized values is zero. We can also see that the standard deviation of that difference decreases in an asymptotically order, and approaches zero when $n \rightarrow \infty$. For instance, if we choose to use $b=1$, the standard deviation of the mean of 1600 pixels is equal to ${{2^{ \lfloor \log _{2} 1 \cdot \sigma \rfloor }} \over {\sqrt {3 \cdot 1600}}}$, which is smaller than ${1 \over 69}\sigma$. In typical integer FITS frames, the standard deviation in this case will usually be smaller than 1, which indicates that the mean of a group of many pixels in the original frame will practically be equal to the mean of those pixels in the compressed$\backslash$decompressed frame. Since analysis of background pixels usually involves very many pixels, the mean will be only negligibly affected by the presented quantization.

This analysis is based on two assumptions mentioned above: 1. The mean of the difference between the original values and the quantized values is zero, and 2. The distribution of the differences is approximately uniform. \citep{7} suggests that this is not necessarily true for quanta sizes greater than $2 \sigma$. Therefore, the algorithm can guarantee the preservation of the mean only for values of $b$ such that $b \leq 1$. 

The same approach also applies to the sum:\newline
Let $S_0$ be sum of $n$ pixels values in the original frame and $S_q$ be the sum of the $n$ quantized pixel values in the compressed $\backslash$ decompressed frame.\newline
$S_0 = {\sum_{i=1}^{n}{C_{i}}}$ \newline
$S_q = {\sum_{i=1}^{n}{(C_{i}-\Delta_{i})}} = {\sum_{i=1}^{n}{C_{i}}} - {\sum_{i=1}^{n}{\Delta_{i}}}$ \newline
$\frac{S_q}{S_0} = \frac{{\sum_{i=1}^{n}{C_{i}}} - {\sum_{i=1}^{n}{\Delta_{i}}}}{\sum_{i=1}^{n}{C_{i}}} = 1 - \frac{\sum_{i=1}^{n}{\Delta_{i}}}{\sum_{i=1}^{n}{C_{i}}}$ \newline
Since the most likely value of $\Delta_i$ is 0, the most likely value of ${\frac{\sum_{i=1}^{n}{\Delta_{i}}}{\sum_{i=1}^{n}{C_i}}}$ is also 0. The standard deviation of the expression ${\frac{\sum_{i=1}^{n}{\Delta_{i}}}{\sum_{i=1}^{n}{C_i}}}$ is: \newline
$stddev({\frac{\sum_{i=1}^{n}{\Delta_i}}{\sum_{i=1}^{n}{C_i}}}) =
\frac{n \cdot stddev(\bar \Delta)}{n \cdot \bar C} = 
{{2^{ \lfloor \log _{2} b \cdot \sigma \rfloor }} \over {\sqrt {3n} \cdot \bar C}}
$ \newline
Where $\bar C=\frac{\sum_{i=1}^{n}{C_i}}{n}$ \newline
For instance, when computing the sum of 1600 quantized pixels when $\bar C = 1000$ and $b=1$, will provide $\frac{S_q}{S_0}=1 - {{2^{ \lfloor \log _{2} 1 \cdot \sigma \rfloor }} \over {\sqrt {3 \cdot 1600} \cdot 1000}} > 1 - \frac{\sigma}{69000}$ \newline

Why not quantize all of the pixels, including the bright pixels of obvious point sources?  The practical risk here is that point sources so quantized might involve a small number of pixels and lead to a large error.  For instance, if we have only 5 bright object pixels which can be relied on for photometry of a certain astronomical object, the standard deviation of the mean of the pixel values is ${2^{ \lfloor \log _{2} b \cdot \sigma \rfloor } \over {\sqrt {3 \cdot 5}}}$, which might be $\sim0.26b\sigma$. Since the pixels of the astronomical objects are usually the most interesting to science, the presented algorithm allows a user to completely preserve their values through the quantization process. However, since signal loss due to noise might be greater than signal loss due to the quantization process, a user might choose to set $d$ to $\infty$ in order to force the algorithm to quantize {\it all} the pixels in the frame. This will increase the compression factor (examples are given in Section 5), but will also result in additional signal loss. Even though the additional signal loss caused by the quantization can be smaller than the signal loss already caused by the background noise, since those pixels are the most valuable for science we chose to allow absolute preservation of their values. This also allows a user to use extreme values of $b$, while still preserving the point spread functions of the sources. Quantizing the brightest pixels can not only change the raw sum of these pixels, but also the point spread function.  Note that the point spread function shapes can be useful for everything from photometry to discerning cosmic-rays. 

Also, in some case, the photometric brightness of a source can be estimated from the brightest source pixels alone, after background subtraction.  Such brightness measures are particularly useful when the background changes significantly and unpredictably over the wings of the PSF.  One project that uses such photometric measures is the Night Sky Live project, which struggles against a ill-behaved sloping background and so records quantities like C1, C5, C9, etc., meaning the level of the brightest pixel, the average of the five brightest pixels, etc.  Preserving the brightest pixel values then specifically enables such photometry schemes.  

Lastly, in cases of sub-pixel point spread functions, when only the single brightest pixel is measured, not preserving the brightest pixel could lead to a  loss of any signal of on the order of $b \sigma$.

Concentrating again on the background, using a large number of pixels reduces the affect of the quantization on the mean, so that the mean of the compressed $\backslash$ decompressed frame should be practically equal to the mean of the original frame. Assuming a normal distribution of the pixel values, the {\it median} of the original frame should be equal to the {\it mean} of the original frame.  Since the mean of the original frame is nearly preserved, the median of the original frame can be equated to the mean of the compressed$\backslash$decompressed frame.  However, since the value of the median pixel is changed, computing the median directly from the compressed$\backslash$decompressed frame produces a different value than the median of the original frame.  Therefore, the median preservation is strongly subject to the assumption of normal distribution of the pixel values. Practically, the number of pixel values involved in the median computation is finite, and can actually be smaller than the quantum size. As the number of pixels involved is smaller, the perturbation to the median is higher. In the worst case scenario, the perturbation can be $b\sigma$. In some cases, this perturbation may not easily avoided, but since $b$ is user defined, a user can know and control the maximum perturbation allowed for the median. Due to that perturbation, users of the compressed$\backslash$decompressed data might need to adjust their photometry algorithms accordingly. In order to avoid computing the background using the median or mode, users might choose, for instance, to use the mean with outlier rejection.

A higher $d$ quantizes more and brighter pixels.  Therefore a lower $d$ makes fainter objects peak above the quantization and hence easier to detect in the compressed $\backslash$ decompressed frame. However, a relatively high $b$ might also contribute to the lossiness of faint objects. For instance, when $b=1$, faint objects with pixel values such that $C_{x,y}-B_{x,y} < \sigma$ might be completely gone from the compressed$\backslash$decompressed frame. Therefore, one using the algorithm should consider her expectations of photometry integrity and set the $b$ and $d$ parameters accordingly. We believe that an advantage of our method is that it provides a guarantee on the lost signal in standard statistical terms.

The local background estimation is particularly efficient when the ratio of the number of point source pixels to background pixels is small.  Additionally, the median filter window size ($s$) used for background computation should be significantly larger than the point spread function of the astronomical sources.  Yet, the median (or mean) is sometimes not an optimal background estimation when the objects are large \citep{11}.  In astronomy, fortunately for our PHOTZIP algorithm, variability is prevalent only among small objects always below the angular size of the point-spread function.

A non-uniform unknown background is best determined locally.  In pictures such as those taken by the Night Sky Live project, the background is highly non-uniform and usually unpredictable, making a local background estimation highly important. The local estimation also simplifies the usage of the algorithm, since one does not have to be aware of the type of the background of the frame when applying the compression. Nevertheless, in cases a uniform background is desired, the $t$ parameter describe in Section 3 can be set for that purpose.

\section{Performance of the PHOTZIP Algorithm}

PHOTZIP should be used along with a multi-purpose lossless compression utility such that the lossless utility is applied after running PHOTZIP. We tested PHOTZIP with two common compression utilities: {\it gzip}, which is based on Ziv \& Lempel's compression algorithm \citep{9}, and {\it bzip2}, which is based on Burrows \& Wheeler's algorithm \citep{17}.  These algorithms, like others, are based on finding and compressing repetitive patterns in the data.  Additionally, quantization of the signal increases redundancy and allows multi-purpose compression algorithm a yet higher compression factor \citep{16}.

The compression factor and the amount of lost signal determine the utility of the algorithm. In order to test the algorithm, we used all-sky 1024 $\times$ 1024 FITS images used by Night Sky Live all-sky monitoring network that deploys CONtinuous CAMeras (CONCAMs).  We also tested FITS images taken by other optical instruments. In all cases, the algorithm provided a significantly better compression factor then all popular multi-purpose compression utilities alone. Table \ref{compression_level} below presents the affect of the $d$, $b$, $s$, $t$ parameters on the performance of the algorithm in terms of compression factor, and compares it to compression factors of common lossless compression programs. The four rightmost columns are the compression factors when using PHOTZIP before applying gzip and bzip2, and when using gzip or bzip2 without using PHOTZIP. The compression factor used here is the amount of data that was compressed as a percentage of the size of the original frame. \newline

\begin{table*}[h]
\begin{center}
\caption{Compression Factor}\label{compression_level}
\begin{tabular}{lcccccccccc}
\hline
\bf File Name& \bf File Size& \bf $d$& \bf $b$& \bf $s$& \bf $t$& \bf photzip& \bf gzip& \bf photzip& \bf bzip2 \hfill\\
& & & & & & \bf + gzip& & \bf + bzip2& \bf \hfill\\
\hline
n3166\_lr.fits& 204 KB& 1& 1& 20& 0& 76.6\%& 50.3\%& 81.9\%& 60.1\%\\
n3166\_lr.fits& 204 KB& 2& 1& 20& 0& 79.7\%& 50.3\%& 86.4\%& 60.1\%\\
n3166\_lr.fits \footnotemark[1]& 204 KB& 1& 2& 20& 0& 80.3\%& 50.3\%& 84.4\%& 60.1\%\\
n3166\_lr.fits \footnotemark[1]& 204 KB& 3& 6& 10& 0& 98.6\%& 50.3\%& 98.6\%& 60.1\%\\
n3166\_lr.fits \footnotemark[1]& 204 KB& 3& 3& 20& 0& 91.7\%& 50.3\%& 92.6\%& 60.1\%\\
n3166\_lr.fits& 204 KB& 3& 1& 20& 0& 82.8\%& 50.3\%& 86.4\%& 60.1\%\\
n3166\_lr.fits& 204 KB& $\infty$& 1& 20& 0& 84.0\%& 50.3\%& 87.9\%& 60.1\%\\
n3166\_lr.fits \footnotemark[1]& 204 KB& $\infty$& 2& 20& 0& 91.8\%& 50.3\%& 92.7\%& 60.1\%\\
n3184\_lj.fits& 204 KB& 1& 1& 20& 0& 75.0\%& 50.5\%& 81.3\%& 55.5\%\\
n3184\_lj.fits \footnotemark[1]& 204 KB& 3& 1& 20& 0& 82.8\%& 50.5\%& 87.1\%& 55.5\%\\
ci040325ut005115p.fits& 2102 KB& 1& 1& 8& 0& 66.7\%& 38.1\%& 73.6\%& 54.1\%\\
ci040325ut005115p.fits& 2102 KB& $\infty$& 1& 8& 0& 76.2\%& 38.1\%& 84.0\%& 54.1\%\\
ci040325ut005115p.fits \footnotemark[1]& 2102 KB& 1& 2& 10& 0& 74.0\%& 38.1\%& 77.3\%& 54.1\%\\
ci040325ut005115p.fits \footnotemark[1]& 2102 KB& 2& 3& 10& 0& 82.6\%& 38.1\%& 86.9\%& 54.1\%\\
ci040325ut005115p.fits \footnotemark[1]& 2102 KB& 2& 3& 10& 0.5& 88.7\%& 38.1\%& 91.1\%& 54.1\%\\
ci040325ut005115p.fits& 2102 KB& 3& 1& 10& 0.5& 78.5\%& 38.1\%& 84.8\%& 54.1\%\\
ci040325ut005115p.fits& 2102 KB& 3& 1& 10& 0& 73.4\%& 38.1\%& 81.8\%& 54.1\%\\
ci040325ut005115p.fits \footnotemark[1]& 2102 KB& 1& 2& 10& 0.5& 81.5\%& 38.1\%& 85.2\%& 54.1\%\\
ci040325ut005115p.fits \footnotemark[1]& 2102 KB& 1& 3& 10& 0.5& 87.4\%& 38.1\%& 89.2\%& 54.1\%\\
ci040325ut005115p.fits \footnotemark[1]& 2102 KB& 1& 3& 10& 0& 79.1\%& 38.1\%& 86.9\%& 54.1\%\\
ci040325ut005115p.fits \footnotemark[1]& 2102 KB& 1& 6& 10& 1.15& 92.1\%& 38.1\%& 93.2\%& 54.1\%\\
ci040325ut005115p.fits \footnotemark[1]& 2102 KB& 3& 6& 10& 1.15& 94.9\%& 38.1\%& 95.1\%& 54.1\%\\
\hline
\end{tabular}
\medskip\\
\end{center}
\end{table*}

Figure 1 depicts the image file ci040325ut005115p.fits, a typical FITS image from the NSL \citep{1} project.  The picture is available in FITS and JPG formats at http://www.NightSkyLive.net. The file n3166\_lj.fits and n3184\_lj.fits were taken from the galaxy catalog at http://www.astro.princeton.edu/\char126frei/catalog.htm.  The later is shown as Figures 2 and 3.

As the maximum allowed absolute difference ($b$) increases, the compression factor is higher, but so is the signal loss.  Since the tested frames contained mostly background, like many astronomical images, the value of $b$ has a substantial affect on the compression factor.  To test the utility of the $b$ parameter, we set $d$ to $\infty$, effectively letting the algorithm quantize {\it all} the pixels and therefore achieving a higher compression factor. 
Although a better compression factor can be achieved, not using the $d$ parameter can increase the bright signal loss as discussed in Section 4.

A higher $t$ parameter increases the compression factor, but this parameter should be used with extra caution since it can also potentially cause signal loss of bright pixels.  The $s$ parameter normally does not have a significant affect on the compression factor, but has some affect on the time required to compress an image.

\section{Conclusions}

When even a single pixel is quantized, photometry can be at least partially compromised.  A background estimated from the mean of surrounding pixels, however, will usually end up closer to the original background. We have shown that using even as little as 1600 pixels for background estimation can reduce the background error in the mean of the quantized pixels to below a single count.  Practically, this means that the background estimation and hence photometric measurements can be protected in the PHOTZIP compression$\backslash$decompression process.

In sum, we present here a simple lossy compression algorithm for astronomical images. The main advantage of this algorithm is that it can preserve the signal of bright pixels, while symmetrically losing signal only from the background.  The criteria for preserving a pixel's value is user-defined (in terms of $\sigma$), so the user can accurately control the compression factor/signal loss trade-off.  In addition, the algorithm guarantees a user-defined maximum absolute difference (also in terms of $\sigma$), so the user can control the amount of lost signal for those areas in the frame that are not preserved.  The algorithm was implemented and tested on unsigned 16-bit integer FITS images, but we believe that the same approach can be applied also to signed integer and floating point images. The algorithm was tested on some typical astronomical 16-bit integer FITS images and appeared to be effective.  PHOTZIP is now in routine use on FITS data taken by the Night Sky Live project.

We thank an anonymous referee for numerous and insightful comments.  LS acknowledges fellowship support from Michigan Technological University.  RJN acknowledges support from the National Science Foundation.

\clearpage

\begin{figure*}
\begin{center}
\includegraphics[scale=1, angle=0]{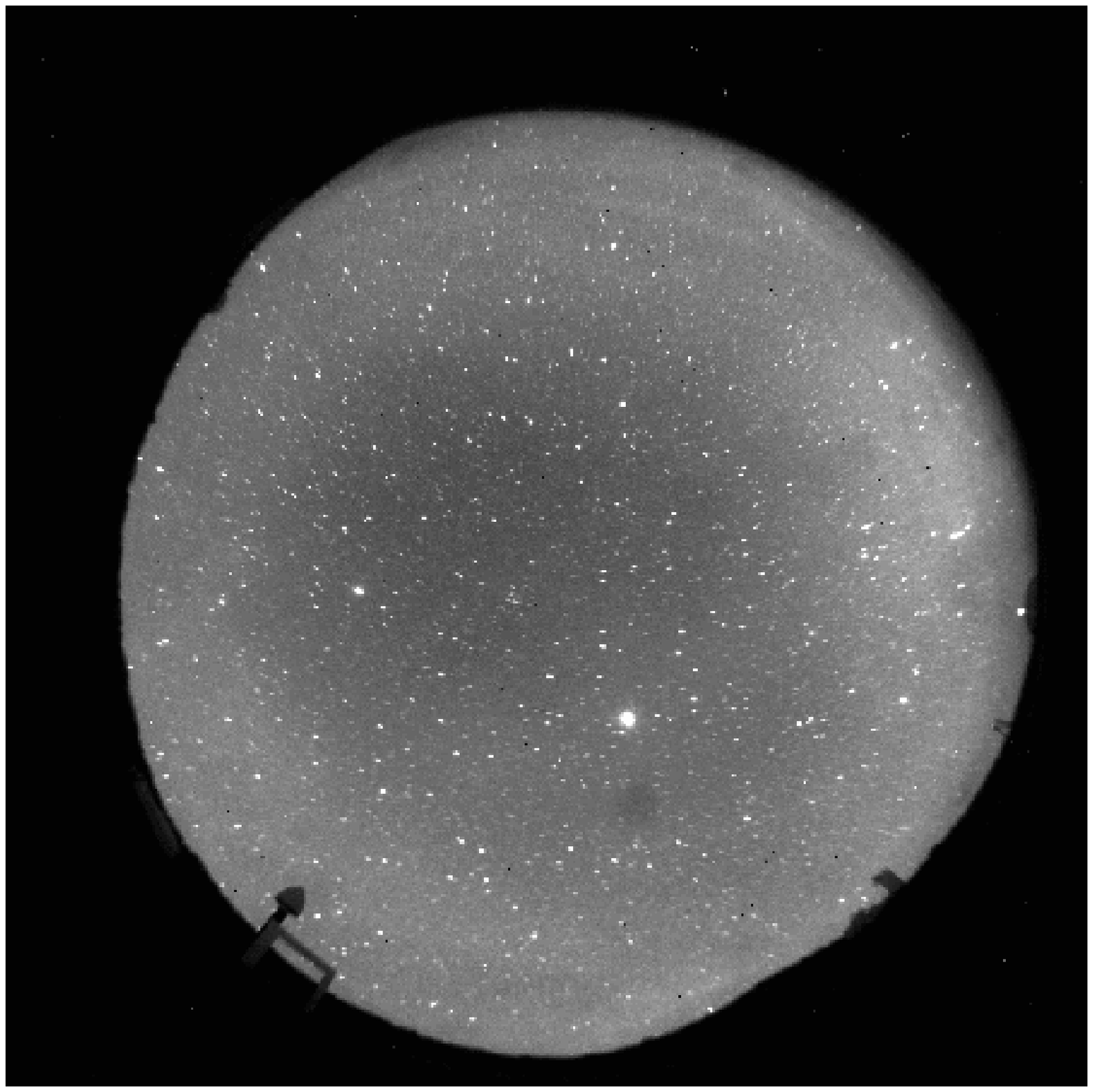}
\caption{ci040325ut005115p.fits: A Night Sky Live all-sky picture. \newline
http://www.NightSkyLive.net}
\label{fig1}
\end{center}
\end{figure*}

\begin{figure*}
\begin{center}
\includegraphics[scale=1, angle=0]{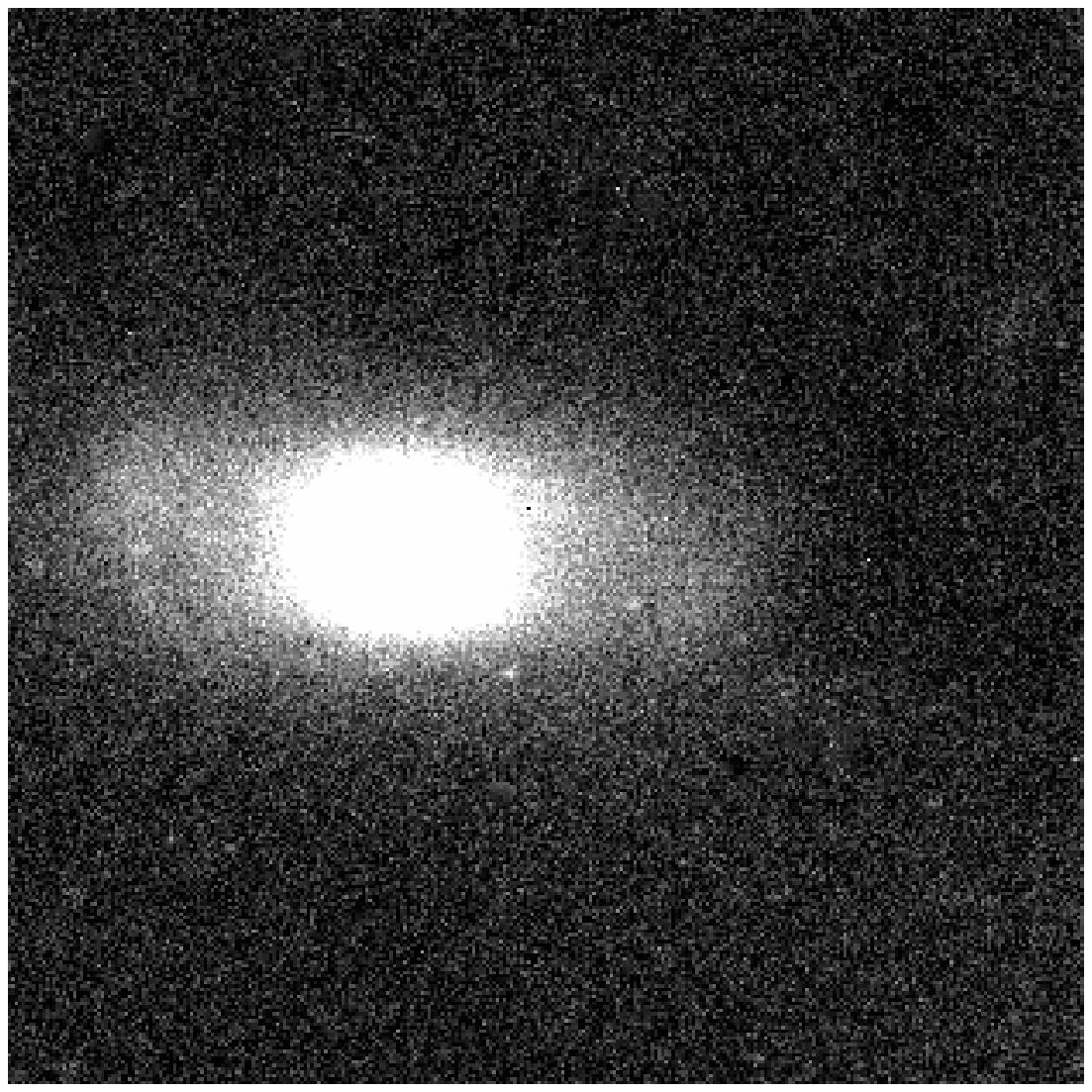}
\caption{n3166\_lr.fits: A picture taken from the {\it galaxy catalog}. \newline http://www.astro.princeton.edu/\char126frei/catalog.htm.}
\label{fig2}
\end{center}
\end{figure*}

\begin{figure*}
\begin{center}
\includegraphics[scale=1, angle=0]{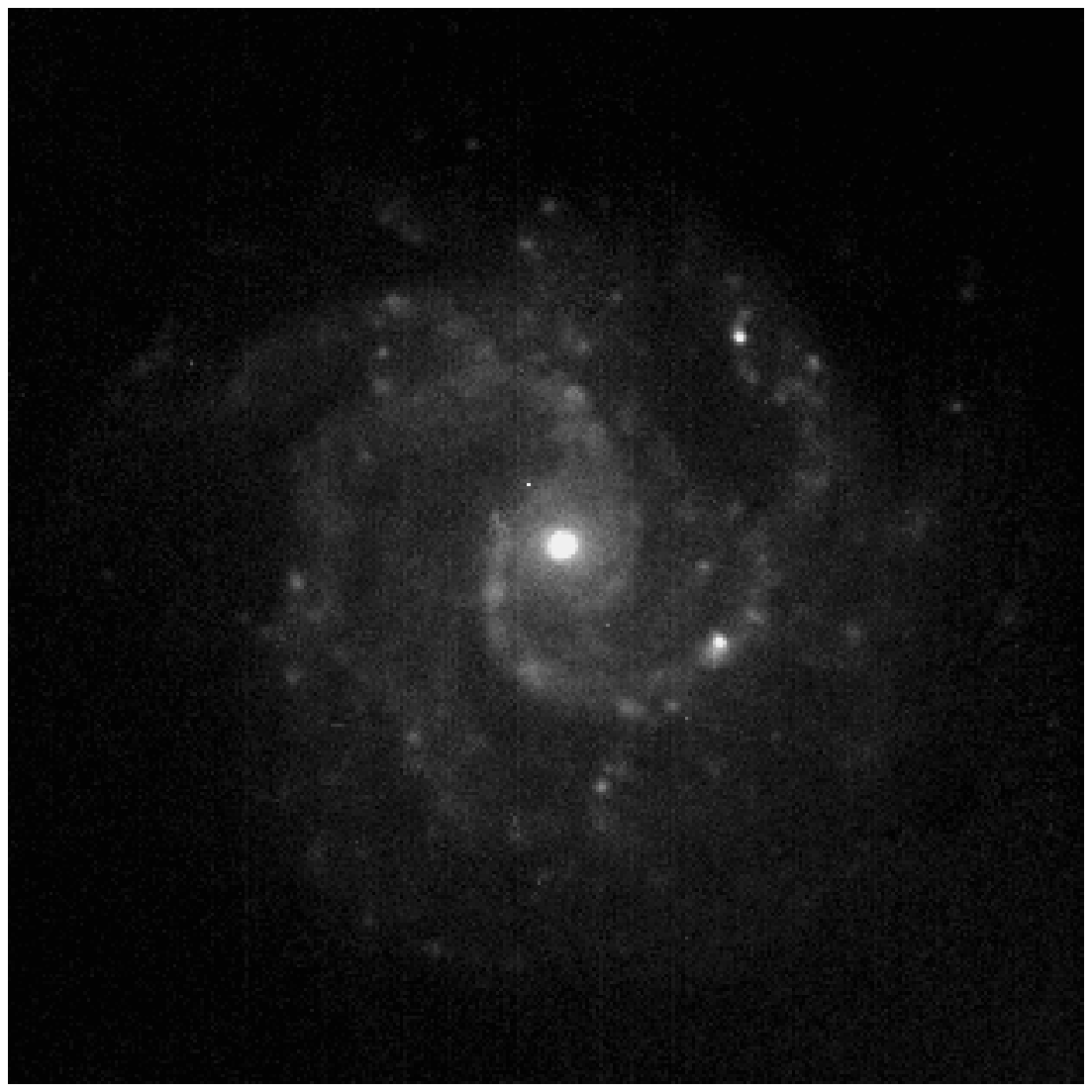}
\caption{n3184\_lj.fits: A picture taken from the {\it galaxy catalog}. \newline 
http://www.astro.princeton.edu/\char126frei/catalog.htm.}
\label{fig3}
\end{center}
\end{figure*}

\end{document}